\documentclass[aps,prb,twocolumn,groupedaddress,draft,showpacs,intlimits,amsmath,amssymb,floats]{revtex4}
\newcommand{\bk}{\textbf{k}}
\newcommand{\bq}{\textbf{q}}
\newcommand{\bp}{\textbf{p}}
\newcommand{\iwm}{i\omega_m}
\newcommand{\iwn}{i\omega_n}
\usepackage{bm}
\usepackage[final]{graphicx}
\usepackage{epsfig}
\begin{document}

\title{Density of States Modulations from Oxygen Phonons in $d-$wave Superconductors: 
Reconciling Angle-Resolved Photoemission Spectroscopy and Scanning Tunneling Microscopy}

\author{S. Johnston$^{1,2}$}
\author{T. P. Devereaux$^{2,3}$}
\affiliation{$^1$Department of Physics and Astronomy, University of Waterloo, Waterloo, Ontario, N2L 3G1, Canada.}
\affiliation{$^2$Stanford Institute for Materials and Energy Science, SLAC National Accelerator Laboratory and 
Stanford University, Stanford, CA 94305, USA}
\affiliation{$^3$Geballe Laboratory for Advanced Materials, Stanford University, Stanford, CA 94305, USA}

\date{\today}

\begin{abstract}
Scanning tunneling microscopy (STM) measurements 
have observed modulations in the density of states (DOS) of a 
number of high-T$_c$ cuprates.  These modulations have been interpreted 
in terms of electron-boson coupling analogous to 
the dispersion ``kinks" observed by angle-resolved 
photoemission spectroscopy (ARPES).  However, a direct a reconciliation 
of the energy scales and features observed by the two probes is presently lacking.  
In this paper we examine the general features of el-boson coupling in a 
$d-$wave superconductor using Eliashberg theory, focusing   
on the structure of the modulations and the 
role of self energy contributions $\lambda_z$ and $\lambda_{\phi}$.  
We identify the features in the DOS that correspond to the gap-shifted 
bosonic mode energies and discuss how the 
structure of the modulations provides information about an underlying 
pairing mechanism and the pairing nature 
of the boson.  We argue that the scenario most 
consistent with the STM data is that of a low-energy boson mode 
renormalizing over a second dominant pairing interaction and we 
identify this low-energy mode as the out-of-phase bond buckling oxygen phonon.
The influence of inelastic damping on the phonon-modulated DOS is 
also examined for the case of Bi$_2$Si$_2$CaCu$_2$O$_{8+\delta}$. 
Using this simplified framework we are able to account for the observed isotope 
shift and anti-correlation between the local gap and mode energies.  
Combined, this work provides a direct reconciliation of the bandstructure 
renormalizations observed by both ARPES and STM in terms of 
coupling to optical oxygen phonons.
\end{abstract}

 \maketitle

\section{Introduction}
In conventional superconductors, the role of electron-phonon (el-ph) interactions 
as the pairing ``glue" was confirmed by the observation of fine-structure in the 
tunneling-derived density of states (DOS). \cite{McMillan, Doug}  
In the cuprates, 
a great deal of effort has been expended in search of similar signatures using 
different probes. \cite{DahmNature2009, KimPRL2003, LanzaraNature2001, 
BogdanovPRL2000,JohnsonPRL2001,KaminskiPRL2001, 
LeeNature2006, Zhu2006, BalatskyPRB2006, PasupathyScience2008, 
ShimPRL2008, ZhaoPRB2007, ZhaoPRL2009, PilgramPRL2006, ZhouPRL2005, IwasawaPRL2008, NormanPRL1997, 
JiangPRB1996, SchachingerPRB2009, SchachingerPRB2008, SchachingerPRB2010,  LeePRB2007, LeePRB2008,
McElroyScience2005, AlldredgeNature2008, Zasadzinski, SandvikPRB2004, tpdPRL2004, Fischer, 
JohnstonACMP, MeevasanaPRL2006, LeePRL2009, ChenPRL2009}  
The observation of a particular mode, 
be it spin or lattice, might be used to infer the nature of the principal pairing 
mediator.  However, to date, there is no consensus on the existence of such a 
mode \cite{Anderson} and if it does exists, whether it is tied to the 
lattice or spin degrees of freedom.  

Electronic renormalizations in the form of dispersion ``kinks" have also been 
observed in the electronic dispersion of the cuprates 
by angle-resolved photoemission spectroscopy (ARPES).\cite{BogdanovPRL2000,JohnsonPRL2001,
KaminskiPRL2001,LanzaraNature2001,KimPRL2003,tpdPRL2004,LeePRB2008,
IwasawaPRL2008,ChenPRL2009, LeePRL2009} 
Occurring at an energy scale $\sim 65-70$ meV in the nodal direction 
($0$,$0$) - ($\pi$,$\pi$) of the Brillouin zone, these kinks have been 
interpreted as coupling to a bosonic mode of either an electronic
\cite{BogdanovPRL2000, JohnsonPRL2001, KaminskiPRL2001, KimPRL2003, 
NormanPRL1997,SchachingerPRB2008,DahmNature2009} 
or lattice origin.\cite{LeePRB2007,LeePRB2008,tpdPRL2004,LanzaraNature2001,JohnstonACMP, 
IwasawaPRL2008,MeevasanaPRL2006,LeePRL2009,ChenPRL2009} 
According to Eliashberg theory, the position of a kink due to 
coupling to a sharp bosonic mode in a $d$-wave superconductor is expected 
to occur at an energy 
of $\Delta_0 + \Omega_0$, where $\Omega_0$ is the energy of the mode and 
$\Delta_0$ is the maximum value of the $d$-wave 
gap\cite{Doug,SandvikPRB2004,LeePRB2008}  With this 
observation, the energy of the mode responsible for the kink has 
been estimated at $\sim 35-40$ meV.  This energy  
coincides with the energies of the spin-resonance mode centered at 
${\bf Q} = (\pi/a,\pi/a)$ observed by neutron 
scattering\cite{Neutron} and 
the out-of-plane ``$B_{1g}$" bond buckling oxygen modes.\cite{tpdPRL2004}  
As a result, there is considerable debate as to the identity of the 
underlying mode, however the observation of multiple mode coupling in 
some cuprates,\cite{MeevasanaPRL2006, ZhouPRL2005, JohnstonACMP} 
as well as recent measurements of an isotope effect for the kink 
energy,\cite{IwasawaPRL2008} provide strong evidence for the phonon interpretation 
of the ARPES renormalizations.  However, regardless of the identity of the 
boson, the $\sim 35-40$ meV energy scale extracted from ARPES data appears to be  
in conflict with the $\sim 52$ meV energy scale extracted from STM by 
Lee {\it et al.}.\cite{LeeNature2006}
Therefore it is an open question as to whether the two experimental probes 
are observing different mode couplings or if they are 
reflecting different manifestations of the same bosonic mode. 

Microscopic inhomogeneities in the local density of states (LDOS)  
which is proportional to the derivative of the tunneling current 
$N(\omega)\propto dI/dV$, as well 
as signatures of coupling to a bosonic mode, have been observed in scanning 
tunneling microscopy (STM) experiments on Bi$_2$Sr$_2$CaCu$_2$O$_{8+\delta}$ 
(Bi-2212). \cite{LeeNature2006, Zhu2006, McElroyScience2005}  
In Ref. \onlinecite{LeeNature2006}, estimates for the local gap size 
are determined from the peak-to-peak distance of the coherence peaks 
while the energy of the bosonic mode is identified as the energy 
position of a peak in $d^2I/dV^2$ measured relative to the energy of the 
superconducting gap. While the positions of 
the superconducting coherence peaks vary at different tip locations, 
estimates for the mode energy are inversely correlated with the local gap 
size and the distribution of mode estimates, centered at $\sim 52$ meV, 
shows a clear isotope shift upon $^{18}$O substitution.  
The energy of the bosonic mode also appears to be immune to doping 
while the spectra changes qualitatively.  These observations 
are inconsistent with a coupling to the spin resonance mode,\cite{Fischer}  
and points to a lattice origin for the mode involving oxygen vibrations.   

Fine structures in the DOS of a number of additional cuprates have been 
reported by other STM and SIS junction tunneling measurements, each producing 
different estimates for mode energies.  A structure similar to that reported for Bi-2212 
has  been reported in Bi$_2$Sr$_2$Ca$_2$Cu$_3$O$_{10+\delta}$ (Bi-2223),\cite{PasupathyScience2008} 
with the energy scale of the mode estimated at $\sim 35$ meV.  It is important to note 
that in Ref. \onlinecite{PasupathyScience2008} the mode energy was associated with the 
position of the minima in the LDOS (root in $d^2I/dV^2$) relative to the local gap size as 
opposed to Ref. \onlinecite{LeeNature2006}, which extracted the mode estimate 
from the peak in $dI^2/dV^2$.  Bicrystal 
grain boundary SIS junction measurements on optimal doped La$_{1.84}$Sr$_{0.16}$CuO$_4$ thin 
films observe modulations in the DOS which correspond well with peaks in the 
neutron derived phonon spectra,\cite{ShimPRL2008} and in agreement with the multiple 
features present in ARPES data.\cite{ZhouPRL2005} 
Examinations of tunneling data on YBa$_2$Cu$_3$O$_{7-\delta}$ (YBCO) 
and the electron doped system Pr$_{0.88}$LaCe$_{0.12}$CuO$_4$ have produced a similar 
correspondence between modulations in the DOS and phonon density of states in 
these materials. \cite{ZhaoPRB2007, ZhaoPRL2009}  
The observation of multiple mode coupling, as well as the fact that the 
spin resonance is well separated in energy from the phonons in the electron 
doped systems provide further evidence that 
the spin resonance mode is unlikely to be the source of these features 
as proposed by Ref. \onlinecite{Fischer}. 
However, the question remains whether 
coupling to the mode can be cast in the usual form for el-ph coupling to 
oxygen modes, as inferred from ARPES \cite{LeePRB2007, tpdPRL2004}, 
or whether the structure in the  
LDOS could be due to phonon-assisted co-tunneling from the tip via 
the apical atom.\cite{PilgramPRL2006}

Co-tunneling via a strong local coupling of electrons in the STM tip 
to the apical atom imparts structure in the form of peaks in the 
$dI/dV$ at an energy scale of $\Delta_0+\Omega$ and multiples of the 
phonon frequency $\Delta_0 + 2\Omega$, $\Delta_0 + 3\Omega$, ... 
even though the coupling of the planar superconducting electrons 
to the apical atom may be weak.\cite{LambePR1968,PilgramPRL2006}  
However, recent observations of charge-ordering structures via Fourier transform 
spectroscopy in Bi-2212, \cite{KohsakaScience2007} which has 
apical oxygen atoms, and Ca$_{2-x}$Na$_x$CaO$_2$Cl$_2$, 
\cite{HanaguriNature2007} which doesn't, empirically indicate that 
no one particular matrix element controls the tunneling pathway 
along the $c$-axis.  If a single pathway were to dominate the 
tunneling process one would expect the charge ordering features 
in these two materials to have qualitatively different structures.  

In light of these 
observations we investigate the role of el-ph coupling between planar 
oxygen vibrations and the electrons of the CuO$_2$ plane.   The aim of 
this paper is twofold.  
First, due to the discrepancies in the structures in $dI/dV$ used 
to extract boson mode energies in previous works, 
we examine the qualitative signatures of el-boson 
coupling in a $d$-wave superconductor.  In doing so, we explicitly 
determining which features in $dI/dV$ (or $d^2I/dV^2$) are 
best identified with the energy of the 
boson mode.  Furthermore, we demonstrate how the qualitative structure in 
$dI/dV$ can be used to determine if the mode coupling is 
the predominant pairing mechanism.  We argue that most 
consistent interpretation of the STM data is that of a low-energy mode 
renormalizing over a dominant high-energy (or instantaneous) pairing interaction.  
In this case, the mode energy should be extracted from a {\it minimum} in 
$N(\omega) \propto dI/dV$.  In light of this finding, the mode energy 
estimate from Ref. \onlinecite{LeeNature2006} is revised to $\sim 35-45$ meV, 
in agreement with Ref. \onlinecite{PasupathyScience2008} and the energy 
scale extracted from ARPES measurements.\cite{LeePRB2008,JohnstonACMP}      
The second goal of this paper is to demonstrate that the modulations in 
$N(\omega)$ can be understood in terms of the same el-ph coupling models  
that have been successful in accounting for the dispersion ``kinks"  
observed ubiquitously in the cuprates by ARPES.  Combined, these calculations 
demonstrate a unified picture of el-ph coupling in the cuprates probed by 
both STM and ARPES.      

The organization of this 
paper is as follows.  In section II the Eliashberg framework 
for calculating the self-energy due to el-boson coupling in a 
$d$-wave superconductor is reviewed.  In section III focus is placed on 
el-boson contributions in multiple momentum channels in order to demonstrate 
the qualitative changes in the self-energy that are expected 
for a mode which couples differently in each of the momentum channels.  
We find that these considerations are critical and can qualitatively change 
the structure of the boson modulations in the DOS,  
altering the feature in $dI/dV$ corresponding to the mode energy.    
This aspect of the calculation has generally been neglected in 
previous Eliashberg treatments, \cite{SchachingerPRB2008,SchachingerPRB2009, 
SchachingerPRB2010,JiangPRB1996} where the boson contribution to 
the single-particle ($\lambda_z$) and anomalous ($\lambda_\phi$) 
self-energies have been taken to be equal to (or proportional to) 
one another over the entire energy range of the boson spectrum.  
For an $s$-wave superconductor $\lambda_z = \lambda_\phi$ but for 
a $d$-wave superconductor $\lambda_z \ne \lambda_\phi$ and usually 
$\lambda_z \gg \lambda_\phi$ in modes derived from the strong Coulomb 
repulsion having strong onsite ($\lambda_z$) and near-neighbor ($\lambda_\phi$) 
interactions.  
Next, having established the role of the symmetry channels, 
we then turn to the qualitative differences in the el-boson structures 
in the infinite and finite band formalisms of Eliashberg theory.  
Since the cuprates have a relatively narrow bandwidth due to the 
Cu 3$d$ character of the $pd$-$\sigma^*$ band, one expects that the latter case 
is the more appropriate formalism for these materials.  
Finally, in section IV we present a model calculation for the el-ph modulated 
LDOS in Bi-2212. Here it is demonstrated that 
the DOS modulations can be reproduced using coupling 
strengths similar to those used in previous ARPES treatments.\cite{tpdPRL2004} 
We then introduce a phenomenological treatment of local damping effects 
reflecting the inhomogeneity observed in Bi-2212 and the associated 
broadened spectral features.\cite{McElroyScience2005} 
This model is able to reproduce the 
anticorrelation between the local superconducting gap estimates and 
local mode estimates.  Finally, the reported isotope shift of the 
distribution mode estimates is naturally captured by the el-ph 
coupling model, as our calculation explicitly show.  
We then end with a brief summary and discussion about the 
implications of this work.

\section{Electron-Boson Coupling}
In Migdal-Eliashberg theory, which neglects crossing diagrams, 
the self-energy due to el-boson coupling is given by 
\begin{eqnarray}\label{Eq:Migdal}\nonumber
\hat{\Sigma}(\bk,i\omega_n) = \frac{1}{N\beta} \sum_{\bq,m}
\int_0^\infty d\nu 
\frac{2\nu\alpha^2F(\bk,\bq,\nu)}{\nu^2+(\omega_m-\omega_n)^2} \times \\
\hat{\tau}_3\hat{G}(\bp,i\omega_m)\hat{\tau}_3
\end{eqnarray} 
where $\omega_m$ $(\omega_n)$ is a Boson (Fermion) 
Matsubara frequency, and $\alpha^2F(\bk,\bq,\nu) = -|g(\bk,\bq)|^2\mathrm{Im}D(\bq,\nu)$ 
is the effective electron-boson spectral function.  
Here, the vertex $g(\bk,\bq)$ gives the strength of scattering of the 
electron from 
state $\bk$ to $\bp = \bk-\bq$ and $D(\bq,i\omega_m)$ is the boson propagator.  

In the superconducting state, the electron propagator   
$\hat{G}(\bk,i\omega_n)$ is given by
\begin{eqnarray}\label{Eq:prop}\nonumber
\hat{G}^{-1}(\bk,i\omega_n) = i\omega_nZ(\bk,i\omega_n)\hat{\tau}_0 
&+& [\epsilon_\bk+\chi(\bk,i\omega_n)]\hat{\tau}_3\\ &+&\phi(\bk,i\omega_n)
\hat{\tau}_1.
\end{eqnarray}
Here, the self-energy has been divided into the canonical form 
\cite{Doug} 
\begin{eqnarray}\label{Eq:Sigma} \nonumber
\hat{\Sigma}(\bk,i\omega_n) = i\omega_n[1 - Z(\bk,i\omega_n)]\hat{\tau}_0 
&+& \chi(\bk,i\omega_n)\hat{\tau}_3 \\
&+&\phi(\bk,i\omega_n)\hat{\tau}_1
\end{eqnarray}
where $i\omega_n[1-Z(\bk,i\omega_n)]$ and $\chi(\bk,i\omega_n)$ are  
the odd and even components of the electron self-energy, respectively, 
and $\phi(\bk,i\omega_n)$ is the anomalous self-energy. Eqs. 
(\ref{Eq:Migdal}) - (\ref{Eq:Sigma}) can now be self-consistently solved 
on the Matsubara frequency axis.  
An analytic expression for the self-energy 
on the real axis can be derived from these equations by explicitly 
performing the analytic continuation 
$i\omega_n \rightarrow \omega + i\delta$, where $\delta > 0$ is an infinitesimal real 
number.  The single-particle DOS is $N(\omega)$ is obtained from the 
electron Green's function $N(\omega) = -2\sum_\bk \mathrm{Im}G_{11}(\bk,\omega)/\pi$.

Marsiglio {\it et al.} \cite{MarsiglioPRB1988} 
have developed an efficient iterative procedure for performing the analytic 
continuation of the el-boson self-energy.   
In this formalism $\hat{\Sigma}(\bk,\omega)$ is obtained by iteratively solving 
\cite{MarsiglioPRB1988,MarsiglioPRB1990} ($\bp = \bk-\bq$) 
 \begin{eqnarray}\label{Eq:Marsiglio}\nonumber 
  \hat{\Sigma}(\bk,\omega)&=&\frac{1}{N\beta}\sum_{\bq}\sum_{m=0}^{\infty} 
  \lambda_0(\bk,\bq,\omega-\iwm) \hat{\tau}_3 \hat{G}(\bp,\iwm)\hat{\tau}_3 \\
 &&+\frac{1}{N}\sum_\bq \int_{-\infty}^\infty d\nu \alpha^2F(\bk,\bq,\nu)\times \\ \nonumber  
 &&[n_b(\nu) + n_f(\nu-\omega)]  
 \hat{\tau}_3\hat{G}(\bp,\omega-\nu)\hat{\tau}_3 
 \end{eqnarray}
where $n_b$ and $n_f$ are the Bose and Fermi factors, respectively, and  
\begin{equation*} 
\lambda_0(\bk,\bq,\omega) = \int_0^\infty d\nu 
\alpha^2F(\bk,\bq,\nu)\frac{2\nu}{\omega^2 - \nu^2}.  
\end{equation*}
In evaluating Eq. (\ref{Eq:Marsiglio}), the solution to the imaginary axis 
equations $\hat{\Sigma}(\bk,\iwn)$ are used as input and $\hat{\Sigma}(\bk,\omega)$ is 
solved for iteratively.\cite{MarsiglioPRB1988} 
 
In large bandwidth systems the momentum sum is typically evaluated by assuming a linear band 
near the Fermi level $\epsilon_\bk = (k-k_F)\cdot v_F$ (with $v_F$ the Fermi 
velocity), replacing the normal state density of states $N(\omega)$ 
with its value at the Fermi level $N(0)$ and extending the limits of the 
energy integral to infinity.\cite{MarsiglioPRB1988}  (Note that in 
this approximation $\chi(\bk,\omega)$ is featureless and only contributes to an overall 
shift of the chemical potential.\cite{Doug}) For brevity we refer to this approximation 
as the ``infinite band" formalism of Eliashberg theory.  

\section{Qualitative Signatures of El-Boson Coupling} 
Before turning to the model calculations, let us discuss the qualitative 
signatures of el-boson coupling in a $d$-wave superconductor.  
We would like to examine coupling in various momentum channels and, to 
this end, the coupling constant can be expanded in terms of Brillouin 
zone harmonics
\begin{equation}
|g(\bk,\bq)|^2 = \sum_{J,J^\prime} Y_J(\bk)|g_{J,J^\prime}|^2
Y_{J^\prime}(\bk-\bq)
\end{equation}
where the sum $J$ runs over the irreducible representations of the point 
group of the crystal.  We assume $g_{J,J^\prime}$ is diagonal in this 
basis and identify $J = 0$ and $J = 2$ with the $s$ and $d_{x^2-y^2}$ 
symmetries.  If one then only admits the gap solution in the $J = 2$ channel, 
the energy and momentum dependence of the self-energies are factorable  
with $Z(\bk,\omega) = Z(\omega)Y_0(\bk)$, 
$\chi(\bk,\omega) = \chi(\omega)Y_0(\bk)$ and 
$\phi(\bk,\omega) = \phi(\omega)Y_J(\bk)$, where $J = 0$,$2$ for each symmetry.  

The overall strength of the el-boson coupling in momentum channel 
$J$ can be parameterized by the dimensionless constant $\lambda_{J}$
\begin{equation}\label{Eq:lambda}
\lambda_J = \int_0^\infty \frac{2d\nu}{\nu}
\frac{\sum_{\bk,\bq}\alpha^2F(\bk,\bq,\nu)
Y_J(\bk)Y_J(\bp)\delta(\epsilon_\bk)\delta(\epsilon_\bp)}
{\sum_\bk Y_J(\bk)^2\delta(\epsilon_\bk)}.
\end{equation}
It is important to note that $\lambda_{J=0} \equiv \lambda_z$ 
($Y_0(\bk) = 1$) characterizes  
the contribution to the el-boson self-energies $Z(\bk,\omega)$ and 
$\chi(\bk,\omega)$ while $\lambda_{J=2} \equiv \lambda_\phi$ 
($Y_{2} = [\cos(k_xa)-\cos(k_ya)]/2$) 
characterizes the contribution to $\phi(\bk,\omega)$ for a 
$d_{x^2-y^2}$ superconductor.  The 
relative values of $\lambda_{z,\phi}$ also determine the 
transition temperature T$_c$, which, in the limit of weak coupling 
is given by
\begin{equation}\label{Eq:Tc}
k_bT_c = 1.13\hbar\Omega_0 \exp\left[-\frac{1+\lambda_z}{\lambda_\phi}\right]. 
\end{equation}
The relative magnitudes of $\lambda_{z,\phi}$ can also affect 
the qualitative signatures of the boson modulations in the density of states.   

\begin{figure}[tl]
 \includegraphics[width=\columnwidth]{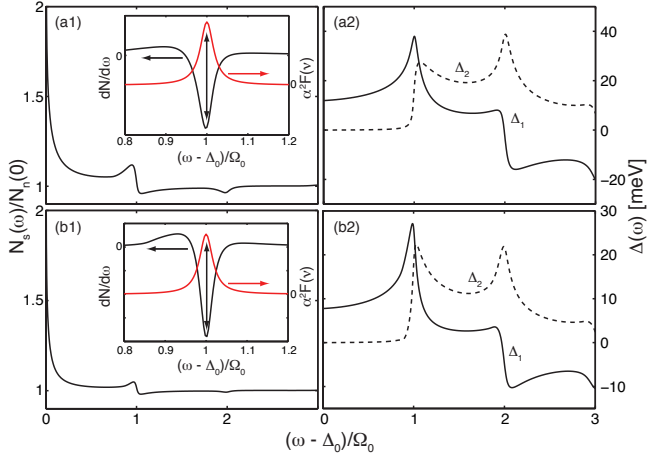}
 \caption{\label{Fig:1} (Color online) 
 (a1), (a2) The DOS $N(\omega)$ and corresponding 
 real and imaginary parts of the gap function $\Delta(\omega)$, 
 respectively, for an $s$-wave superconductor coupled to a single 
 bosonic mode.  (b1), (b2) $N(\omega)$ and $\Delta(\omega)$ for a 
 $d$-wave superconductor with $\lambda_z = 1.6$ and 
 $\lambda_\phi = 0.8\lambda_z$.  
 The insets of panels (a1),(b1),   
 show $dN/d\omega$ and $\alpha^2F(\nu+\Delta_0)$ in order to 
 highlight the correspondence between features in the DOS and 
 peaks in the boson spectrum.  The secondary structure in 
 $\Delta(\omega)$ at $\omega = \Delta_0 + 2\Omega_0$ are due to 
 self-energy effects due to two-phonon processes.  
 }
\end{figure}

In the superconducting state, neglecting the role bandstructure, 
the DOS $N_s(\omega)$ can be written as 
\begin{equation}\label{Eq:DOS}
\frac{N_s(\omega)}{N_n(0)} = \mathrm{Re} \left\langle 
\frac{\omega}{\sqrt{\omega^2 - \Delta^2(\bk,\omega)}} \right\rangle
\end{equation} 
where $N_n(0)$ is the density of states at the Fermi level and 
$\Delta(\bk,\omega) = \phi(\bk,\omega)/Z(\bk,\omega) = 
\Delta_1(\bk,\omega)+i\Delta_2(\bk,\omega)$ is the complex momentum-dependent 
gap function and $\langle...\rangle$ denotes an average over the Fermi 
surface.\cite{Doug}  
In what follows, the notation $\Delta_0$ is introduced 
for the maximum value of the 
superconducting gap on the Fermi surface, which is defined by the  
value of the gap function at the gap edge 
$\Delta_0 = \Delta(\bk_f^\mathrm{AN},\omega=\Delta_0)$, where 
$\bk_f^{\mathrm{AN}}$ is the Fermi momentum along the zone face 
(0,0)-($\pi$,$0$). 

For $\omega \gg \Delta_0$, Eq. (\ref{Eq:DOS}) can be expanded 
yielding \cite{Doug}
\begin{equation}\label{Eq:Nexpand}
\frac{N_s(\omega)}{N_n(0)} = 1+\frac{1}{2\omega^2}
\langle \Delta_1^2(\bk,\omega) - \Delta_2^2(\bk,\omega)\rangle. 
\end{equation} 
From Eq. (\ref{Eq:Nexpand}) it is clear that the phononic substructure is 
given by the frequency dependence of the $\Delta(\bk,\omega)$, which is 
obtained by evaluating Eq. \ref{Eq:Marsiglio}.  
In Figs. \ref{Fig:1}a1-a2 we plot $N_s(\omega)$ and the corresponding $\Delta(\omega)$, 
respectively, for an isotropic $s$-wave superconductor. 
Here, we have assumed a single bosonic mode 
with a Lorentzian spectral density, characterized by a 
half-width at half-maximum of $\Gamma_{b} = 1$ meV and centered at $\Omega_0 = 52$ meV  
(inset of panel Fig. \ref{Fig:1}a1).  The overall coupling strength 
has been set to $\lambda_z = 0.8$.

As $\omega \rightarrow \Delta_0 + \Omega_0$, the real part of $\Delta(\omega)$ 
begins to rise producing an enhancement in the DOS for $\omega \lesssim 
\Delta_0 + \Omega_0$.  At $\Delta_0 + \Omega_0$ the real part begins to 
drop while the imaginary part experiences a sudden rise due to Kramers-Kronig 
consistency.  This results in a rapid suppression in the DOS at this energy 
scale, which drives the DOS below its bare value.  As a result, the energy 
scale $\Delta_0 + \Omega_0$ manifests as a shoulder in $N(\omega)$ or a minimum in 
$dN/d\omega$, as shown in the inset of Fig. \ref{Fig:1}a1.  This is the classic 
McMillan-Rowell signature, similar to that observed in Pb, where phonons are 
solely responsible for pairing. \cite{Doug, McMillan} 
In the case of a $d$-wave superconductor   
the situation is nearly identical when $\lambda_z = \lambda_\phi = 0.8$, 
and the magnitude of the gap function is comparable to that obtained 
for the $s$-wave case with a similar value of $\lambda_z$ (not shown).  
Again, since the 
contribution from the boson is equal in the two channels, the boson energy 
scale manifests as a shoulder on the low-energy side of the modulation.  
This signature qualitatively unchanged when $\lambda_z > \lambda_\phi$, 
as shown in Figs. \ref{Fig:1}b1,b2 for $\lambda_z = 1.6$, $\lambda_\phi=0.8\lambda_z$, 
with the overall magnitude of $\Delta(\omega)$ reduced by a factor of 2.  
The reduction in $\Delta(\omega)$ illustrates the importance of the 
relative magnitudes of $\lambda_z$ and $\lambda_\phi$.   
 
We now consider the case of a low-energy 
boson which renormalizes over a second mode, higher in energy and  
dominant in its contribution to pairing.  We associate this 
high energy mode with spin fluctuations with a broad Lorentzian spectral density 
centered at $\Omega_{sf} \sim 2J = 260$ meV and with $\Gamma_{sf} = 30$ meV.  The 
total coupling to this mode is taken to be $\lambda_{sf,z} = \lambda_{sf,\phi} = 1.6$.  
For the low-energy boson we take $\Gamma_{b} = 0.5$ meV,  $\lambda_z = 1.6$ and 
$\lambda_{\phi} = 0$.  

\begin{figure}[tr]
 \includegraphics[width=0.60\columnwidth]{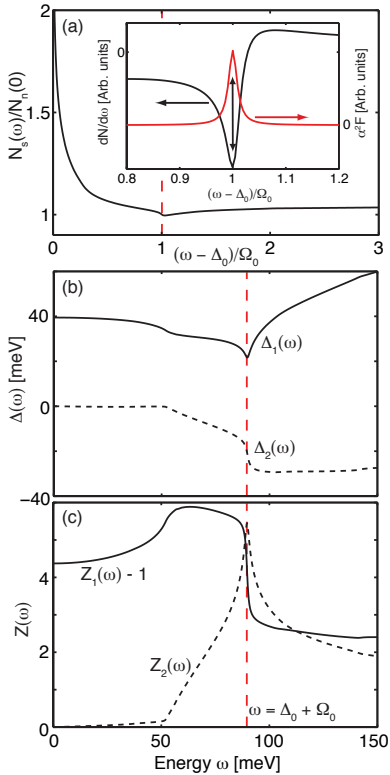}
 \caption{\label{Fig:2mode} (Color online) 
 (a) $N(\omega)$ for a $d$-wave superconductor 
 where the $\Omega \sim 52$ meV mode 
 ($\lambda_{b,z} = 1.6$, $\lambda_{b,\phi} = 0$) 
 renormalizes over a dominate mode 
 associated with spin fluctuations with $\Omega_{sf} = 260$ meV 
 ($\lambda_{sf,z} = \lambda_{sf,\phi} = 1.6$).  Inset: 
 $dN/d\omega$ and $\alpha^2F(\nu-\Delta_0)$ highlighting the correspondence 
 between the low-energy mode and structure in $N(\omega)$.  (b), (c) 
 $\Delta(\omega)$ and $Z(\omega)$ for this case.  The red dashed lines 
 indicate the energy of $\Delta_0 + \Omega_0$ where $\Omega_0$ is the 
 energy of the center of the low-energy spectra density (inset, panel (a)).}
\end{figure}

The results for the two-mode calculation are shown in Fig. \ref{Fig:2mode}. 
The structure of the renormalizations in this case 
differs considerably and  
the shoulder on the low energy side of the renormalization is significantly 
less pronounced.  
Here, the sharper spectral density and increased 
value of $\lambda_z$ for the low-energy mode are required in order to 
accentuate the weak low-energy feature.  Without this increased coupling the low-energy 
renormalizations are difficult to resolve.   
In the Eliashberg formalism, a high energy boson produces a gap function whose 
real part is relatively frequency independent for energies on the order of 
lower energy mode $\Omega_{b}$, while an instantaneous pairing 
interaction produces a frequency independent gap function up to an energy scale set by 
the Coulomb interaction.\cite{MaierPRL2008}  Therefore, assuming a 
frequency independent pair field $\phi_0$, modulated by el-boson coupling, we write
\begin{equation*}
\frac{\phi(\omega)}{Z(\omega)} = \frac{\phi_0+\delta\phi(\omega)}{Z_0+\delta Z(\omega)}
= \frac{\phi_0}{Z_0}\frac{1+\delta\phi/\phi_0}{1+\delta Z/Z_0}
\end{equation*}
where $\delta\phi$ and $\delta Z$ are the el-boson contributions to the self-energy. 
The DOS (for $\omega \gg \Delta_0$) can then be written as 
\begin{equation}
\frac{N_s(\omega)}{N_f} = 1 + \frac{\Delta_0^2}{2\omega^2}
\left(1-\frac{\delta\phi(\omega)}{\phi_0} - \frac{\delta Z(\omega)}{Z_0} \right).
\end{equation}
If the el-boson contribution to pairing is small, as is the case in Fig. \ref{Fig:2mode}, 
$\delta\phi$ can be neglected and one sees 
that the fine structure tracks the structure of $\delta Z(\omega)$ (Fig. \ref{Fig:2mode}c).  
In Fig. \ref{Fig:2mode}b, this is seen as the suppression of $\Delta_1(\omega)$ as 
$\omega \rightarrow \Delta_0+\Omega_0$ and results in a dip structure in 
$N(\omega)$ (Fig. \ref{Fig:2mode}a) with no pronounced shoulder on the 
low-energy side of the renormalization. 
We also note that the boson energy scale remains as the minima 
in $dN/d\omega$ (inset of Fig. \ref{Fig:2mode}a).  

\begin{figure}[tr]
 \includegraphics[width=\columnwidth]{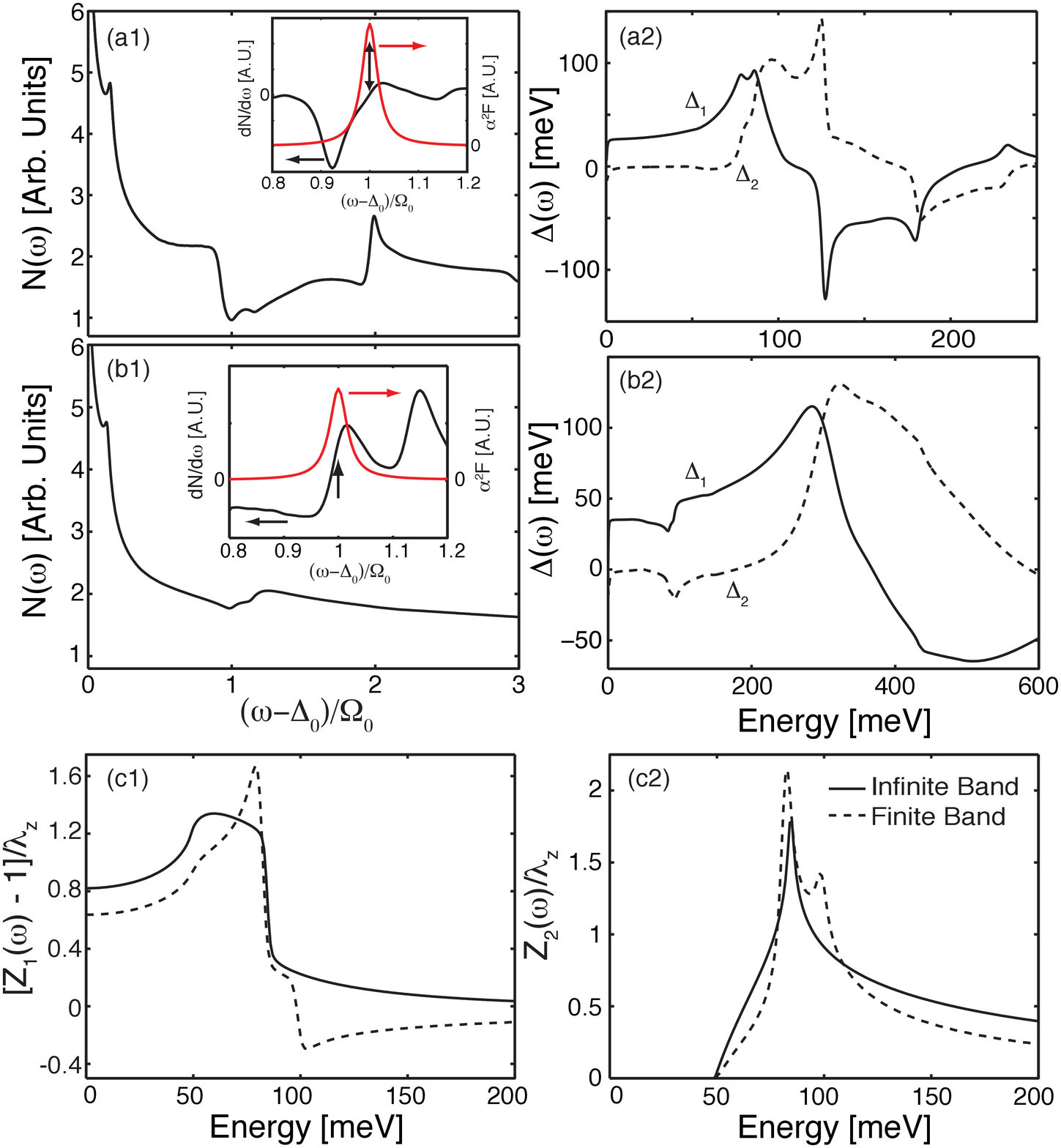}
 \caption{\label{Fig:2} (Color online)  
 The structures in the DOS $N(\omega)$ and gap function $\Delta(\omega)$ 
 when the structure of the bare band has been retained.  
 (a1),(a2) $N(\omega)$ and $\Delta(\omega)$, respectively, 
 for a $d$-wave superconductor coupled to a single low-energy mode which 
 provides all of the pairing.  (b1),(b2) $N(\omega)$ and $\Delta(\omega)$ 
 for the two mode case analogous to the case shown in Figs. \ref{Fig:1}c1,c2.  
 The insets of panels (a1) and (b1) show $dN/d\omega$ and the low energy 
 component of $\alpha^2F(\nu+\Delta_0)$ in order to show the correspondence 
 between structure in $N(\omega)$ and peaks in $\alpha^2F(\nu)$.
 (c1), (c2) The real and imaginary parts of $Z(\omega)$ calculated in the 
 infinite- (solid) and finite-band (dashed) Eliashberg formalisms. Here, a single 
 iteration of the Eliashberg equations has been assumed with coupling to 
 an Einstein mode centered at $\Omega = 52$ meV and assuming a $d$-wave gap 
 with $\Delta_0 = 35$ meV.  
}
\end{figure}

Experimentally, the modulations in the tunneling spectra appear as 
a dip-hump structure with no pronounced shoulder on the low-energy 
side of the modulations.\cite{LeeNature2006, PasupathyScience2008}  
We therefore conclude that, within validity of Eliashberg 
theory, the bosonic mode responsible for the LDOS renormalizations cannot be the 
sole contributor to pairing in the cuprates otherwise a pronounced shoulder 
on the low-energy side of the renormalization would be present.  
More likely, the low-energy boson renormalizes 
over the dominant source of pairing.  This approach has been invoked previously 
in order to account for the ARPES kinks since the value of 
$\lambda_{z} \sim 0.3-0.5$ needed to reproduce the kink is too small to 
provide enough pairing to account for the large gap 
values.\cite{SandvikPRB2004, LeePRB2007}

Although the two mode picture can 
account for the lack of shoulder feature in the 
LDOS in the infinite band formalism, the hump structure on the 
high-energy side of the renormalizations is absent.  This is due to the 
approximations inherent to the formalism used so far, which neglect 
the role of both bandstructure and the narrow bandwidth of the system.  
In Fig. \ref{Fig:2} solutions to the Eliashberg equations 
given by Eq. (\ref{Eq:Marsiglio}), which retain the full $\bk$-dependence 
of the band structure, are shown. Again, two cases are considered in analogy to 
Figs. \ref{Fig:1} and \ref{Fig:2mode}: a single mode with 
$\lambda_z = 1.9 = \lambda_\phi$, $\Gamma = 1$ meV, and $\Omega_0 = 52$ meV, 
and a two-mode case with the modes parameterized by 
$\Omega_{b,sf} = 52$, $300$ meV $\Gamma_{b,sf} = 1$, $30$ meV, 
$\lambda_{z,sf}=\lambda_{\phi,sf} = 1.7$, $\lambda_{z,b} = 0.52$ and 
$\lambda_{\phi,b} = 0$. 
In order to model a realistic bandstructure for the low energy dispersion 
have assumed a 5-parameter tight-binding model obtained from fits to 
the low-energy dispersion observed by ARPES.\cite{NormanPRB1995}  
Fig. \ref{Fig:2}a1, a2, show $N(\omega)$ and $\Delta(\omega)$, 
respectively, for the single low-energy mode which is pairing, analogous 
to Fig. \ref{Fig:1}, while Fig. \ref{Fig:2}b1,b2 shows the results 
for the two mode calculation, analogous to Fig. \ref{Fig:2mode}.   

The first observation is that the shoulder feature remains when a single  
mode contributes significantly to pairing, while a pronounced dip-hump 
structure is produced when the mode renormalizes over another dominant 
mechanism.  In the latter case the hump is more pronounced due to the additional 
self-energy contributions when scattering to the large DOS near the van Hove.  
In the cuprates the energy of the phonons (and spin 
resonance mode) lie close in energy to both the superconducting gap $\Delta_0$ and  
the van Hove energy $\epsilon(0,\pi/a)$.  Because of this near degeneracy of energy 
scales there is an overall enhancement of the self-energy due to the 
increased density of states to which the bosons can couple  
anti-nodal region. This degeneracy has also been noted in studies on the 
temperature dependence of the el-ph self-energy observed by 
ARPES.\cite{tpdPRL2004, LeePRB2008}  To illustrate this 
point, in Fig. \ref{Fig:2}c1,c2 compares the real and imaginary parts of 
$Z(\omega)$ in the two formalisms.  (In order to compare comparable 
cases $Z(\omega)$ as been calculated assuming a single iteration of the 
Eliashberg equations and the results have been normalized by the value of 
$\lambda_z$.) When the bandstructure has been retained $Z(\omega)$ (as well 
as $\chi$ and $\phi$) develops additional structure and $Z_1(\omega)$ 
becomes negative at energies larger than $E(0,\pi) + \Omega_0$, 
where $E^2(0,\pi) = \epsilon^2(0,\pi) + \Delta^2(0,\pi)$ is the energy of 
the quasiparticle at the van Hove singularity, resulting 
in the hump structure observed in $N(\omega)$.  If the bandstructure is 
neglected $Z_1(\omega)$ remains positive and smoothly 
approaches zero as $\omega \rightarrow \infty$.  One can therefore conclude 
that the structure in the underlying band can contribute to the structure in 
the self-energy and must be considered in realistic 
treatments of narrow bandwidth systems such as the cuprates.   

The second observation made of Fig. \ref{Fig:2} is that   
the correspondence between the minima in $dN/d\omega$ and peaks in $\alpha^2F$ 
no longer holds for a $d$-wave superconductor with a finite 
bandwidth (insets of Figs. \ref{Fig:2}a1 and 
b1). For the single-mode model the peak in $\alpha^2F(\nu)$ corresponds to the minimum 
in $N(\omega)$ or the root in $dN/d\omega$ while for the two-mode model 
this energy scale does not correspond to {\it any} feature in $dN/d\omega$.  For the 
choice of $\alpha^2F$ and $\epsilon_\bk$ used here, the energy scale $\Delta_0 + \Omega_0$ 
is located between the root and maximum in $dN/d\omega$ and either energy scale 
gives a reasonable estimate for the mode energy.  

\begin{figure}[t]
 \includegraphics[width=0.8\columnwidth]{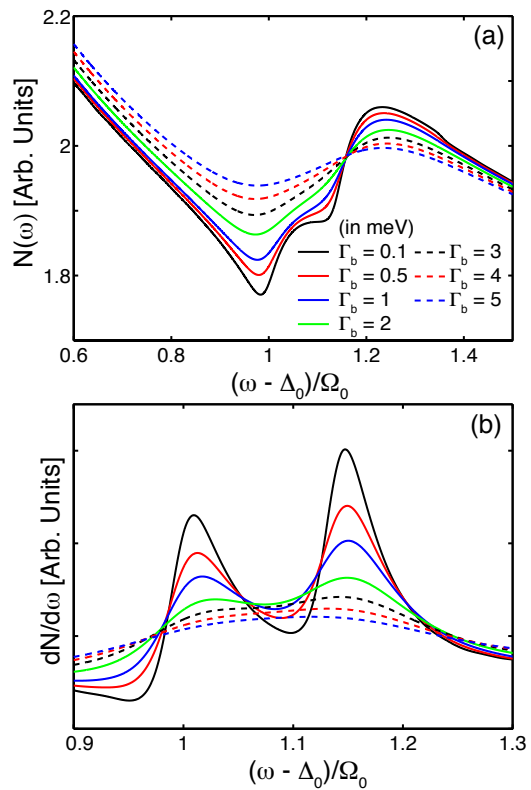}
 \caption{\label{Fig:dndw_vs_gamma} (Color online) 
 (a) $N(\omega)$ and (b) $dN/d\omega$ 
 for $\omega$ in the neighbourhood of the low-energy boson renormalizations 
 for various values of the low-energy mode's spectral width $\Gamma_b$ 
 (in meV).  }
\end{figure}

We also note the presence 
of secondary features in the DOS at an energy $E(0,\pi) + \Omega_0$ 
which is absent when momentum structure of the band 
is neglected.  The presence of this feature can complicate the identification of the 
mode energy if the boson spectral density is broad enough.   
In Fig. \ref{Fig:dndw_vs_gamma} $N(\omega)$ and $dN/d\omega$ are plotted 
for the two-mode model for various values of the low-energy mode's spectral width 
$\Gamma_b$.  Here, we have kept the details of the high-energy mode fixed to the 
same values used in Fig. \ref{Fig:2} and adjusted the strength of $\lambda_{b}$ 
such that the total value of $\lambda_{z,\phi}$ are fixed.  As the value of 
$\Gamma_b$ is increased, the width of the structure in $N(\omega)$ is 
increased and eventually becomes broad enough that the double dip structure 
merges into a single dip-hump feature.  As a result, the peaks in 
$dN/d\omega$ merge into a single peak whose maximum occurs between 
$\Omega_0 + \Delta_0$ and $\Omega_0 + E(0,\pi)$.   The maximum in 
$dN/d\omega$ is very sensitive to the width of the boson 
spectrum as well as the energy of the van Hove and is therefore 
an unreliable indicator of the boson mode energy.  However, the position of the 
root in $dN/d\omega$ on the low energy side of the renormalizations is much 
less sensitive to these factors.  Although this feature slightly 
underestimates the mode energy, it does appear to be a robust indicator of the 
mode energy.  

The qualitative change in the phonon fine-structure is the central result 
of this paper.  It reconciles the discrepancy in the $\sim 52$ meV scale 
observed in STM \cite{LeeNature2006} and the $\sim 70$ meV ``kink" 
observed by photoemission in Bi-2212.\cite{tpdPRL2004}  In Ref. 
\onlinecite{LeeNature2006} the maxima in $d^2I/dV^2 \propto dN/d\omega$ 
was taken for the mode estimate, corresponding to the shoulder of the 
dip-hump structure of the DOS.  However, as shown above, the energy 
scale of the mode is mode accurately given by the minimum (the dip) 
in $N(\omega)$ (or a root in $d^2I/dV^2$), this choice results in an 
overestimate of the mode energy.  A closer examination of Fig. 1 of Ref. 
\onlinecite{LeeNature2006} reveals that the minima in the DOS is $\sim 15$-$20$ 
meV lower in energy.  This brings the mode estimate in line with the energy 
of the $B_{1g}$ modes invoked to explain the kink in the nodal region 
of superconducting Bi-2212.  Indeed, Ref. \onlinecite{PasupathyScience2008} 
has tracked the minimum in the related Bi-2223 system, which also has 
strong coupling to the $B_{1g}$ modes, and obtained a mode energy of 
35 meV, consistent with our findings. 
The qualitative difference in structure also provides a pathway to 
experimentally distinguish between fine structure due to co-tunneling effects, 
where a maximum in 
$N(\Omega_0 + \Delta_0) \propto dI/dV|_{\omega = \Omega_0 + \Delta_0}$ occurs, and 
intrinsic el-boson coupling where a minimum is expected if the mode is not 
dominating pairing, or a shoulder is expected if the mode is contributing 
heavily to pairing.  

\section{Considerations for Bi-2212}
We now turn to a model calculation for Bi-2212.  Here we consider coupling 
to the out-of-phase Cu-O bond buckling $B_{1g}$ phonon branch, previously 
invoked to explain the renormalization in the bandstructure of 
Bi-2212 observed by ARPES.\cite{tpdPRL2004} 
Since we are now concerned with how the phonons renormalize over a dominant 
interaction we consider only a single iteration of the Eliashberg equations 
and treat the phonon mode as a dispersionless Einstein mode with $\Omega_0 = 36$ 
meV.  This treatment is identical to that used in previous works examining 
the dispersion kinks observed by ARPES.\cite{tpdPRL2004,LeePRB2007,LeePRB2008} 
For simplicity, we expand the coupling 
constant as $|g(\bk,\bq)|^2 = g_z^2 + g_\phi^2Y_d(\bk)Y_d(\bp)$ and set 
$g_{z,\phi}$ such that $\lambda_{z,\phi} = 0.31$, $0.1$, comparable to the 
values obtained in previous works.\cite{tpdPRL2004}   
After analytic continuation, the 
zero-temperature expressions for the imaginary parts of the self-energies are:
\begin{eqnarray} 
\omega Z_2(\bk,\omega)&=&\frac{\pi}{2N}\sum_\bp |g(\bk,\bq)|^2\delta(E_\bp + \Omega_0 - \omega)
\\\nonumber
\chi_2(\bk,\omega)&=&-\frac{\pi}{2N}\sum_\bp |g(\bk,\bq)|^2
\frac{\epsilon_\bp}{E_\bp}\delta(E_\bp + \Omega_0 - \omega)
\\\nonumber
\phi_2(\bk,\omega)&=&\frac{\pi}{2N}\sum_\bp |g(\bk,\bq)|^2\frac{\Delta_\bp}{E_\bp}
\delta(E_\bp + \Omega_0 - \omega)
\end{eqnarray}
which are evaluated for $\omega > 0$.  Here, $E(\bk) = \sqrt{\epsilon^2(\bk)+\Delta^2(\bk)}$.  
The real-part of the self-energies are obtained via the Kramers-Kronig relations.  
In these calculations the superconducting gap $\Delta_0$ is taken as an 
input parameter and we supplement the real part of $\phi$ in order to maintain 
the value of the gap at the gap edge.\cite{SandvikPRB2004,LeePRB2007}   
Finally, an intrinsic damping $\Gamma=5$ meV, independent of $\omega$ and $\bk$, is 
added to the imaginary part of $Z(\bk,\omega)$.  

\begin{figure}[tr]
 \includegraphics[width=\columnwidth]{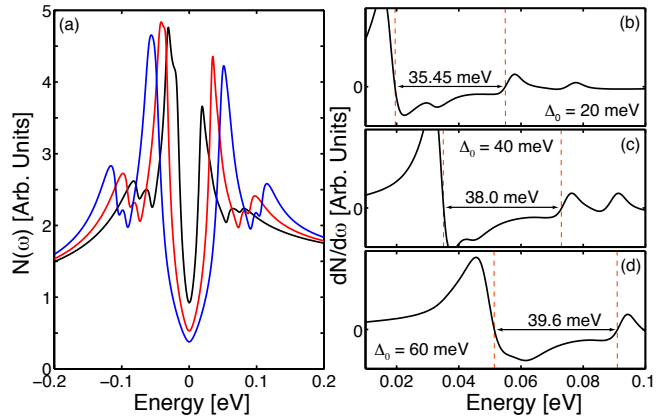}
 \caption{\label{Fig:3} (Color online) 
 (a) The density of states calculated for coupling 
 to the $B_{1g}$ phonon branch at $T=0$.  Each spectra is calculated using 
 a 5-parameter tight-binding bandstucture and a $d$-wave gap with 
 $\Delta_0 = 20$ (black), 40 (red) and 60 (blue) meV.  (b)-(d) 
 $dN/d\omega$ for the indicated gap size.  The red dashed lines indicate 
 the position of the roots corresponding to estimates for 
 $\Delta_0$ and $\Delta_0 + \Omega_0$, respectively.}
\end{figure}

The DOS results are presented in Fig. \ref{Fig:3}, where we have chosen three 
different values of $\Delta_0$ to mimic the variation in the LDOS observed 
in different regions of an optimally doped sample.  The calculated 
DOS for the three gap values all show a clear dip in the spectra around 
$\Delta_0+\Omega_0$ as well as the secondary structure at 
$E(0,\pi) + \Omega_0$ due to the van Hove singularity.  We also note that 
the features associated with the phonon renormalizations are more pronounced 
due to the use of a sharp Einstein mode.    
Estimates are obtained empirically for the mode energy and gap size from the 
calculated DOS via the gap referencing procedure used in Ref. 
\onlinecite{McElroyScience2005}.  
In order to avoid complications associated with the van Hove singularity in 
determining the gap magnitude on the occupied side ($\omega < 0$) 
of the spectra we work on the hole side ($\omega > 0$) and identify the 
energy $\Delta_0+\Omega_0$ with the root in $dN/d\omega$.     
When the gap referencing procedure is applied to the data, the resulting 
gap estimate is equal to the quadrature addition of the gap on the 
Fermi surface $\Delta_0$ and the damping term $\Gamma$.  The empirically 
determined mode energy is therefore underestimated since the 
effective mode position is $\Omega_0 + \Delta_0 - \sqrt{\Delta_0^2 + \Gamma^2}$. 
Since we have used a constant value of $\Gamma$ for each value of $\Delta_0$, 
the small gap data has a larger ratio of $\Gamma/\Delta_0$, and thus 
the extracted mode energy systematically deviated from $\Omega_0$.

A constant value of $\Gamma$ does not capture the local inhomogeneity 
of the parameters entering into the DOS itself.  Specifically, we now modify 
the magnitude of the el-ph coupling and inelastic damping $\Gamma$ included 
in both the spectral function and evaluation of the self-energy, as a function 
of gap size.  We n{\"a}ively associate the larger gap with ``underdoped" 
regions which, due to the reduction in screening of the el-ph interaction, 
leads to an increase in the relative strength of the coupling with gap size.  
At the same time, damping effects are taken to increase together with the gap size 
to mimic the crossover to smeared gap structures found in the large gap regions.  
This is modeled as an increasing ratio of $\Gamma/\Delta_0$ in the large gap region.  
The DOS was then recalculated for input gap values ranging from 15 to 50 meV.  

\begin{figure}[t]
 \includegraphics[width=0.8\columnwidth]{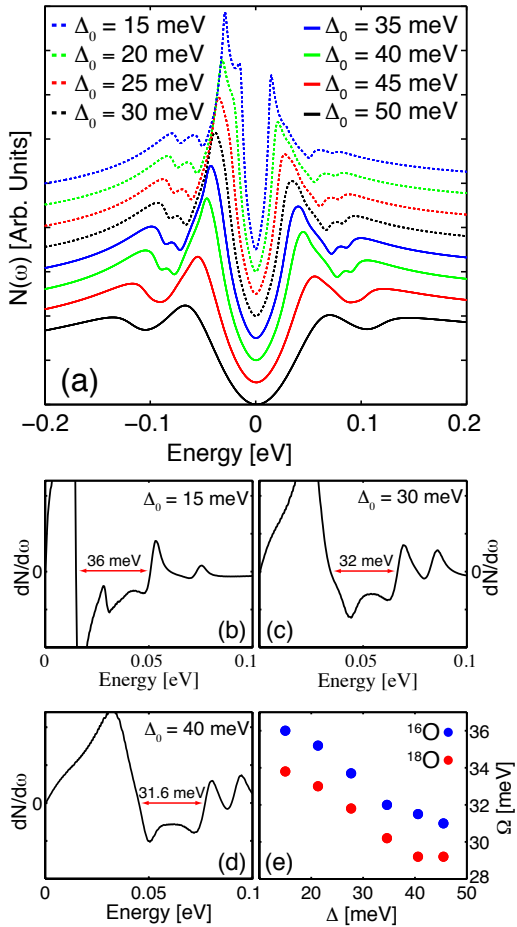}
 \caption{\label{Fig:4} (Color online) 
 A waterfall plot of $N(\omega)$ calculated for 
 doping values spanning the large gap to small gap regions.  Each DOS 
 was calculated for coupling the same mode used in Fig. \ref{Fig:3}.  
 The gap values indicated in the legend denote the input values of 
 $\Delta_0$.  (b),(c),(d) $dN/d\omega$ for selected DOS presented in 
 panel (a) for the hole side ($\omega > 0$) of the spectrum.  The 
 red arrows indicate the roots used to estimate the energies 
 $\Delta_0$ and $\Delta_0 + \Omega_0$.  (e) The mode energy estimate 
 obtained from the position of the local minima relative to the 
 coherence peak. The blue data points correspond to data for $^{16}$O 
 simulations while the red data points correspond to $^{18}$O simulations.}
\end{figure}

The new DOS spectra obtained are presented in Fig. \ref{Fig:4}a.  For 
small gap inputs one can see sharp coherence peaks followed by a well defined 
dip-hump structure associated with the el-ph coupling.  For larger gaps, 
the associated larger damping smears the coherence peaks and they are washed 
away for $\Delta_0 = 50$ meV.  The gap and phonon energy scales can now 
be extracted from $dN/d\omega$, which are shown for selected spectra 
in Figs. \ref{Fig:4}b-d.  The resulting correlation between the extracted 
$\Delta_0$ and $\Omega_0$ are shown Fig. \ref{Fig:4}e (blue dots).  The 
anti-correlation which emerges between the two energies stems from the 
progressive underestimation of the mode energy as the gap size and $\Gamma$ 
are increased.

In order to model the isotope effect, these calculations were repeated 
with adjustments appropriate for the replacement of $^{16}$O with 
$^{18}$O. Specifically, this includes a shift in the phonon frequency by a 
factor of $\sqrt{M_{16}/M_{18}}$ and a decrease in the overall coupling by 
a factor $(M_{16}/M_{18})^{1/4}$. (This substitution leaves the values of 
$\lambda_{z,\phi}$ unchanged.)  The red data points of Fig. \ref{Fig:4}e show the 
correlation between the estimate for $\Omega$ and $\Delta_0$ obtained 
for $^{18}$O upon repetition of the previous calculations.  In both 
cases, the anti-correlation persists and a clear isotope shift can be 
seen, which is on the order of that observed and that 
one would expect based on the known shift in the phonon energy.  
The overall agreement with the experimental data is good, and the anti-correlation 
can be accounted for relatively well by incorporating damping effects into a 
simple el-ph picture.   

\section{Conclusions}
We have examined the signatures of el-boson coupling in a $d$-wave superconductor 
when the boson mode does not provide the dominate pairing interaction.  
The resulting fine structure for the modulations in the DOS are 
qualitatively different than those found in the conventional $s$-wave 
superconductors.  We found that the manifestation of bosonic energy scales in the 
LDOS depends on an interplay between the strength of the el-boson coupling in 
each momentum channel, as well as the details of the underlying bandstructure.  
In order to reproduce the spectra observed experimentally by STM we found that 
both the full structure of the band, as well as the differing contributions of 
$\lambda_z$ and $\lambda_\phi$ for the boson mode in question have to be accounded for.  
Taking these considerations into account, the 52 meV scale reported in 
Ref. \onlinecite{LeeNature2006} is found to be an overestimate of the mode 
energy and instead, the local minima in $N(\omega)$ should be considered.  
The mode energy extracted from the minima of the  
spectra reported in Ref. \onlinecite{LeeNature2006}  
agrees well with the energy of the $B_{1g}$ phonon modes as well as a more 
recent measurements reported for Bi-2223.
\cite{PasupathyScience2008}  This fact reconciles the scales observed 
by STM with those observed in numerous ARPES experiments.  
With the revised estimate for the mode energy, we then calculated 
the LDOS including coupling to the $B_{1g}$ phonon modes.  Using 
a simple consideration for the local intrinsic damping we found that 
the model is able to reproduce both the structure of the renormalizations 
and the observed anti-correlation between the extracted values of $\Omega_0$ and 
$\Delta_0$. The success of the el-ph model in accounting for 
the data obtained by both probes provides further evidence in support of the 
phonon interpretation of the electronic renormalizations observed now in both 
ARPES and STM experiments.  

\acknowledgments
The authors thank E. A. Nowadnick, I. Vishik, B. Moritz, 
W. S. Lee, Z.-X. Shen, J.-H. Lee, J. C. Davis, J.-X. Zhu, 
A. V. Balatsky, P. Hirschfeld, N. Nagaosa, J. Zaanen and D. J. Scalapino 
for many useful discussions. 
This work was supported by the US Department of Energy, Office of 
Basic Energy Sciences under contract no DE-AC02-76SF00515. 
The computational work was made possible by resources of the 
facilities of the Shared Hierarchical Academic Research Computing 
Network (SHARCNET). S. J. would like to acknowledge finacial support 
from NSERC and SHARCNET. We would also like to acknowledge the 
A. von Humboldt Foundation and the Pacific Institute for Theoretical Physics.

\end{document}